\documentclass[aps,preprintnumbers,amsmath,amssymb,latexsym,array,enumerate,letter,twocolumn,superscriptaddress]{revtex4-1}

\usepackage{graphicx}
\usepackage{dcolumn}
\usepackage{bm}
\usepackage{epsfig}
\usepackage{gensymb}
\usepackage[dvipsnames]{xcolor}
\usepackage{mathrsfs}  
\usepackage{amsmath}
\usepackage{amssymb}
\usepackage{graphicx,bm}
\usepackage{slashed}
\usepackage[makeroom]{cancel}
\usepackage[colorlinks = true,
            linkcolor = blue,
            urlcolor  = blue,
            citecolor = blue,
            anchorcolor = blue]{hyperref}
\usepackage{color}

 \usepackage{titlesec}
   \titleformat{\section}[runin]
   {\normalfont\bfseries}{\quad\thesection.}{0.5em}{} [.]
  \titlespacing*{\section}{0pt}{0.2\baselineskip}{\baselineskip}

\def\fun#1#2{\lower3.6pt\vbox{\baselineskip0pt\lineskip.9pt
  \ialign{$\mathsurround=0pt#1\hfil##\hfil$\crcr#2\crcr\sim\crcr}}}

\input epsf

\usepackage{color}

\newcommand{\miniemri}[0]{mini-EMRI\xspace}
\newcommand{\miniemris}[0]{mini-EMRIs\xspace}
\newcommand{\comment}[1]{}
\usepackage{xspace}
\def\gw{GW\xspace}
\def\gwh{GW\xspace}
\def\gws{GWs\xspace}

\def\cwh{continuous-wave\xspace}

\def\cws{continuous waves\xspace}

\def\ecos{ECOs\xspace}
\def\eco{ECO\xspace}
\def\pbhs{PBHs\xspace}
\def\pbh{PBH\xspace}
\newcommand{\TFFT}{T_{\rm FFT}}
\newcommand{\Tobs}{T_{\rm obs}}

\def\erfc{\mathrm{erfc}}

\begin{document}

\preprint{}

\title{Searching for Mini Extreme Mass Ratio Inspirals with Gravitational-Wave Detectors}
\author{Huai-Ke Guo}
\email{huaike.guo@utah.edu}
\affiliation{Department of Physics and Astronomy, University of Utah, Salt Lake City, UT 84112, USA} 

\author{Andrew Miller}
\email{andrew.miller@uclouvain.be }
\affiliation{Université catholique de Louvain, B-1348 Louvain-la-Neuve, Belgium}

\begin{abstract}
A compact object with a mass $\mathcal{O}(1 \sim 1000) M_{\odot}$, such as a black hole of stellar or primordial origin or a neutron star, and a much lighter 
exotic compact object with a subsolar mass could form a non-standard mini extreme mass ratio inspiral (EMRI) and emit gravitational waves within the frequency band of ground-based gravitational-wave detectors. These systems are extremely interesting because detecting them would definitively point to new physics. 
We study the capability of using 
LIGO/Virgo 
to search for mini-EMRIs and find that a large class of exotic compact objects can be probed at current and design sensitivities using a method based on the Hough Transform that tracks quasi power-law signals during the inspiral phase of the mini-EMRI system. 
\end{abstract}


\maketitle







\noindent{\bfseries Introduction.} 
{The direct detection of binary black hole, neutron star and neutron star/black hole mergers by the LIGO/Virgo collaborations \cite{Abbott:2016blz,Abbott:2016nmj,Abbott:2017vtc} has opened the era of gravitational wave (GW) and multi-messenger astronomy. Though these sources were expected, the masses of some of them have surprised us \cite{LIGOScientific:2021usb}, most of all, the existence of $\mathcal{O}(100M_\odot)$ black holes, which cannot be explained by traditional stellar evolution models. Such departures from our current understanding of the universe motivate the study of compact objects that differ from the canonical ones predicted by stellar evolution (white dwarfs, neutron stars, and black holes). Though these so-called exotic compact objects (\ecos), e.g. primordial black holes (\pbhs) \cite{Hawking:1975vcx}, boson stars~\cite{Liebling:2012fv} and quark stars \cite{Weber:2004kj}, have not been detected, their existences could hint at new physics \cite{clesse2018seven}, and help explain puzzling cosmological observations \cite{Planck:2015fie,Ali-Haimoud:2016mbv}. If \ecos are dark, they would constitute a macroscopic dark matter species, in contrast to other hypothetical dark matter candidates, such as weakly interacting massive particles (WIMPs) \cite{Goodman:1984dc}, or ultralight dark matter such as axions \cite{Peccei:1977hh} or dark photons \cite{pierce2019dark}. In particular, the low-spinning black holes detected by LIGO/Virgo indicate that they could have primordial origins \cite{clesse2018seven}, and therefore that \pbhs constitute all or a fraction of dark matter \cite{Carr:2019kxo}.  }

\begin{figure*}[t]
  \includegraphics[width=\textwidth]{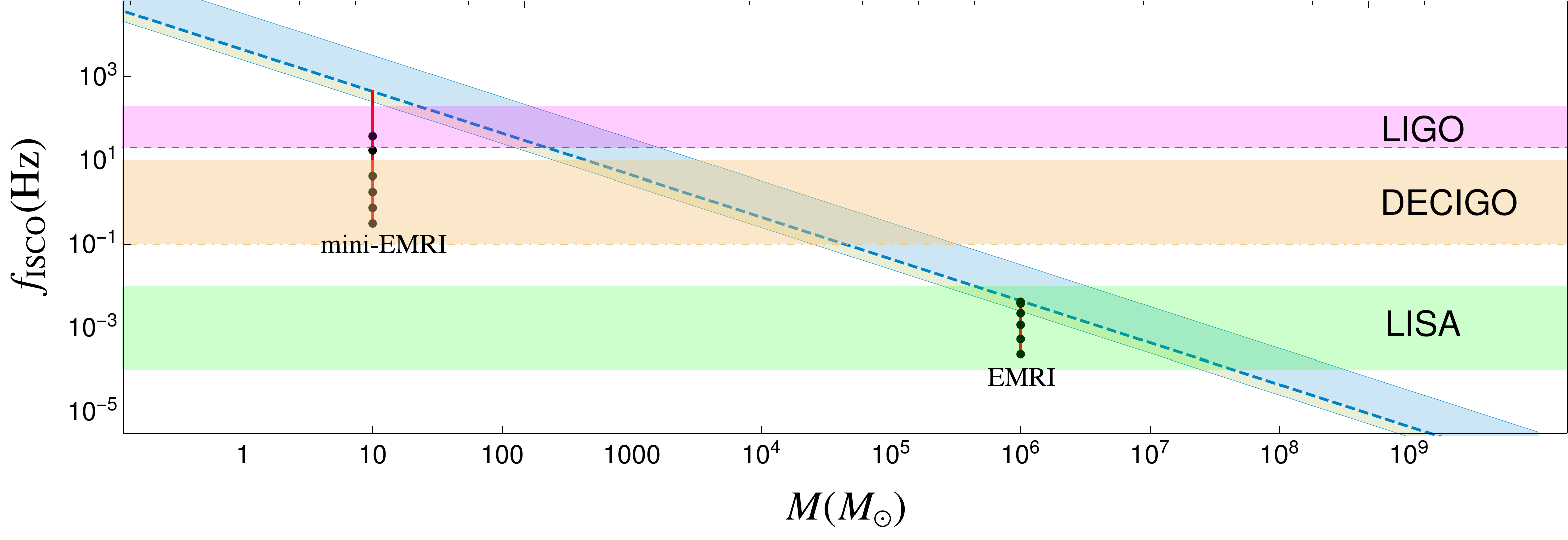}
  \caption{
    \label{fig:fISCO}
    The frequency at the innermost stable circular orbit $f_{\text{ISCO}}$ for an EMRI system as a function of the 
    mass of the heavier object. The tilted blue band is the frequency at ISCO of the massive compact object where its
    spin $a$ varies from $-1$ (lower boundary) to $1$ (upper boundary) with the dashed line denoting $a=0$. 
    The horizontal bands are the frequency regions to which several GW detectors are sensitive. 
    The vertical red lines denote the evolution of the frequency for representative standard EMRI (right) and a \miniemri (left), both with mass ratio $10^{-5}$,
    as the time remaining to ISCO decreases, with the black dots denoting representative time remaining: 1 day, 200 hours, 1 year, 10 years, 100 years 
    and 1000 years from top to bottom.
    }
\end{figure*}

At the moment, however, only limited efforts have been made to detect \ecos by the \gwh community. Searches for sub-solar mass black holes (of $\mathcal{O}(0.1M_\odot)$) have been carried out \cite{Phukon:2021cus,LIGOScientific:2019kan,Nitz:2021vqh}, resulting in constraints on the fraction of dark matter that PBHs could compose.  Scalar \cite{Brito:2017zvb}, vector \cite{baryakhtar2017black,siemonsen2020gravitational}, or tensor \cite{Brito:2020lup} dark matter clouds around black holes have also been theorized, and constraints on boson/black hole mass pairs have been placed using the non-observation of quasi-monochromatic \gwh signals across the whole sky \cite{palomba2019direct,LIGOScientific:2021jlr}. Furthermore, a method has been proposed to search for planetary-mass \pbhs, with masses of $\mathcal{O}(10^{-7}-10^{-2})M_\odot$ \cite{Miller:2020kmv}, and constraints have been placed on the existence of such objects in binary systems \cite{Miller:2021knj,LIGOScientific:2022pjk}. 

However, in order to probe the existence of a variety of \ecos, and cover an extensive portion of the mass parameter space in which \ecos could lie, a more systematic approach is needed. Binary systems composed of lighter \ecos , with masses $\leq \mathcal{O}(0.1M_\odot)$, would emit long-lived \gws in the inspiral portion of their lives \cite{Miller:2020kmv}. Here, we consider a highly \emph{asymmetric} mass ratio between the two compact objects, inspired by the possibility of future space-based \gwh interferometers, such as DECIGO \cite{Kawamura:2020pcg},   LISA~\cite{LISA:2017pwj,Babak:2017tow}, Taiji~\cite{Hu:2017mde,Ruan:2018tsw,Taiji-1} and Tianqin~\cite{TianQin:2015yph,Luo:2020bls,TianQin:2020hid}, to detect extreme mass ratio inspirals (EMRIs). In ground-based detectors, ordinary compact objects would be bound to much lighter \ecos, which we call ``\miniemri'' systems. Depending on the mass ratio of the two compact objects, \miniemris could last anywhere from $\mathcal{O}(\rm hours-days)-\mathcal{O(\rm years)}$ in the frequency band of ground-based \gwh detectors, which would allow signal-to-noise ratio to accumulate over time, as is expected for EMRIs in space-based detectors. A stochastic \gwh background for the kinds of systems considered in this paper could also exist, but is estimated to be very weak \cite{Cui:2021hlu}, motivating the need to consider \emph{individual} \miniemri systems.

The durations of these signals, and the presence of non-stationary noise and gaps in the data, imply computational challenges for traditional matched-filtering algorithms that search for binary neutron star mergers and sub-solar mass and stellar-mass binary black hole mergers \cite{Blelly:2021oim,Dey:2021dem,Sachdev:2019vvd}. Furthermore, \miniemris could last in the detector band for timescales compatible with those expected from \cws from isolated neutron stars \cite{riles2017recent,sieniawska2019continuous}, boson clouds around black holes \cite{DAntonio:2018sff,isi2019directed,LIGOScientific:2021jlr,Sun:2019mqb}, and quasi-monochromatic signals arising from dark matter interactions with \gwh detectors \cite{PhysRevLett.121.061102,guo2019searching,Miller:2020vsl,LIGOScientific:2021odm,Miller:2022wxu}. Hence, methods that have typically been used in these searches can also be applied to detect signals from inspiraling \miniemris.

In this \emph{letter}, we show that a \miniemri system formed by an ordinary or exotic compact object and a 
much lighter ECO can be detected in ground-based \gwh interferometer data.
The current observing run of Advanced LIGO/Virgo can already be used to probe a large region in the parameter space of \ecos, while future detectors will be able to provide even broader coverage.
We also describe a new way to search for these \ecos in the much smaller mass region using the Hough Transform.

\noindent{\bfseries Mini Extreme Mass Ratio Inspiral.}
An EMRI system, in its standard definition~\cite{Babak:2017tow}, consists of a supermassive black hole in the galactic center 
with a mass in the range $10^6 \sim 10^9 M_\odot$, and an inspiraling ordinary compact object such as a black 
hole of astrophysical origin, a neutron star or a white dwarf, with a typical mass of $1\sim 10M_\odot$~\cite{Babak:2017tow}, and
thus a mass ratio $\lesssim 10^{-5}$. 
Non-standard EMRIs could have a much lighter \eco~\cite{Guo:2017njn,Guo:2019sns}, or a superheavy boson star replacing the role of the supermassive black hole~\cite{Torres:2000dw}.
Such EMRI systems are expected to form in the inner parsec region of the galaxies due to complicated stellar dynamics under the influence of the gravitational 
potential of supermassive objects (see, e.g.,~\cite{Alexander:2017rvg,Alexander:2005jz}).

A \miniemri system, instead, is defined here to consist of one object (with mass $M$) much lighter than a supermassive black hole, 
and another one even lighter (with mass $m \ll M$). 
While $M$ can take any value much smaller than that of the supermassive black hole, we restrict to, as an example, searches for a special class of
\miniemris that could be detected by LIGO/Virgo. 
The maximally achievable \gwh frequency from such a \miniemri system, assuming for simplicity a circular orbit, occurs at the innermost stable circular orbit (ISCO)~\cite{Bardeen:1972fi} ,
\begin{eqnarray}
f_{\text{ISCO}} = 4.4 \text{kHz} \left(\frac{1 M_{\odot}}{M}\right) \left(\frac{n}{2}\right) g(a) ,
\end{eqnarray}where $a$ is the dimensionless spin of the heavier component, $n$ is the harmonic number, with $n=2$ being the dominant contributor to \gwh emission, 
and $g$ is a monotonically increasing function of $a$, normalized such that
$g(0)=1$, with $g(-1)\approx 0.57$ and $g(1) \approx 7.35$.
Fig.~\ref{fig:fISCO} shows the band of $f_{\text{ISCO}}$ as a function of $M$ by varying also $a$. 
For the signal to be within the LIGO/Virgo band, we need $M \lesssim \mathcal{O}(1000) M_{\odot}$, and with a mass ratio of $10^{-5}$, 
which requires $m \lesssim \mathcal{O}(10^{-2}) M_{\odot}$. 

An advantage of such \miniemri searches is that the  \gwh amplitude and the signal-to-noise ratio 
generally increase with the chirp mass $M_c \equiv (m M)^{3/5}/(m+M)^{1/5}$; thus, for a given $M_c$, the \miniemri search can probe a much smaller $m$ than what a search for comparable sub-solar mass
binaries (with both masses being $\bar{m}$) can achieve, as 
\begin{eqnarray}
\frac{m}{\bar{m}} \approx 0.8 \left(\frac{\bar{m}}{M}\right)^{2/3} \ll 1 .
\end{eqnarray}This implies that with a larger $M$, we can carve into a much deeper portion of the subsolar mass regime. 
Currently, the most massive black hole detected by LIGO/Virgo was GW190521, in which two merging black holes left behind an intermediate-mass black hole with a mass $142^{+28}_{-16} M_{\odot}$~\cite{LIGOScientific:2020iuh}.
These larger $M$ compact objects could serve as good targets to search for \miniemri \gwh signals, since much lighter ECOs could orbit around them. 
 Furthermore, heavier \ecos would allow us to probe black holes whose mass falls into the traditional intermediate-mass regime.
 While the heavier mass could be an arbitrary ECO, to focus on the search for sub-solar ECOs, we concentrate here on \miniemris with a heavier compact mass, but allow an arbitrary compactness $C$, the dimensionless mass radius ratio, for its lighter partner, such that the parameters 
characterizing the system are $m$, $M$, $a$ and $C$.

The most studied ECO is the PBH, and binaries of PBHs could form in the early universe, either under the torques of nearby PBHs or
density fluctuations~\cite{Nakamura:1997sm,Ioka:1998nz,Ali-Haimoud:2017rtz}, in which \miniemris could arise for a PBH population with an extended mass spectrum~\cite{Kocsis:2017yty,Raidal:2017mfl,Chen:2018czv,Raidal:2018bbj}.  
A \pbh binary could also form through capture in a dense \pbh halo, and the details of this clustering of \pbhs determine the merging rates~\cite{Bird:2016dcv,Clesse:2016vqa}. Additionally, if a stellar-mass black hole passes by a \pbh, it could capture it and form a binary \cite{Cui:2021hlu}. 
Finally, \pbh binaries could be formed by capture in the galactic center \cite{Miller:2020kmv}.
Despite the many competing explanations for what constitutes an \eco, we propose in this work a way to 
probe the existence of \emph{any} sub-solar mass \eco. We are therefore mostly sensitive to the mass $m$ 
and compactness $C$ of the \eco, but are agnostic to how that \eco formed. 


\begin{figure}[t]
  \includegraphics[width=0.7\columnwidth]{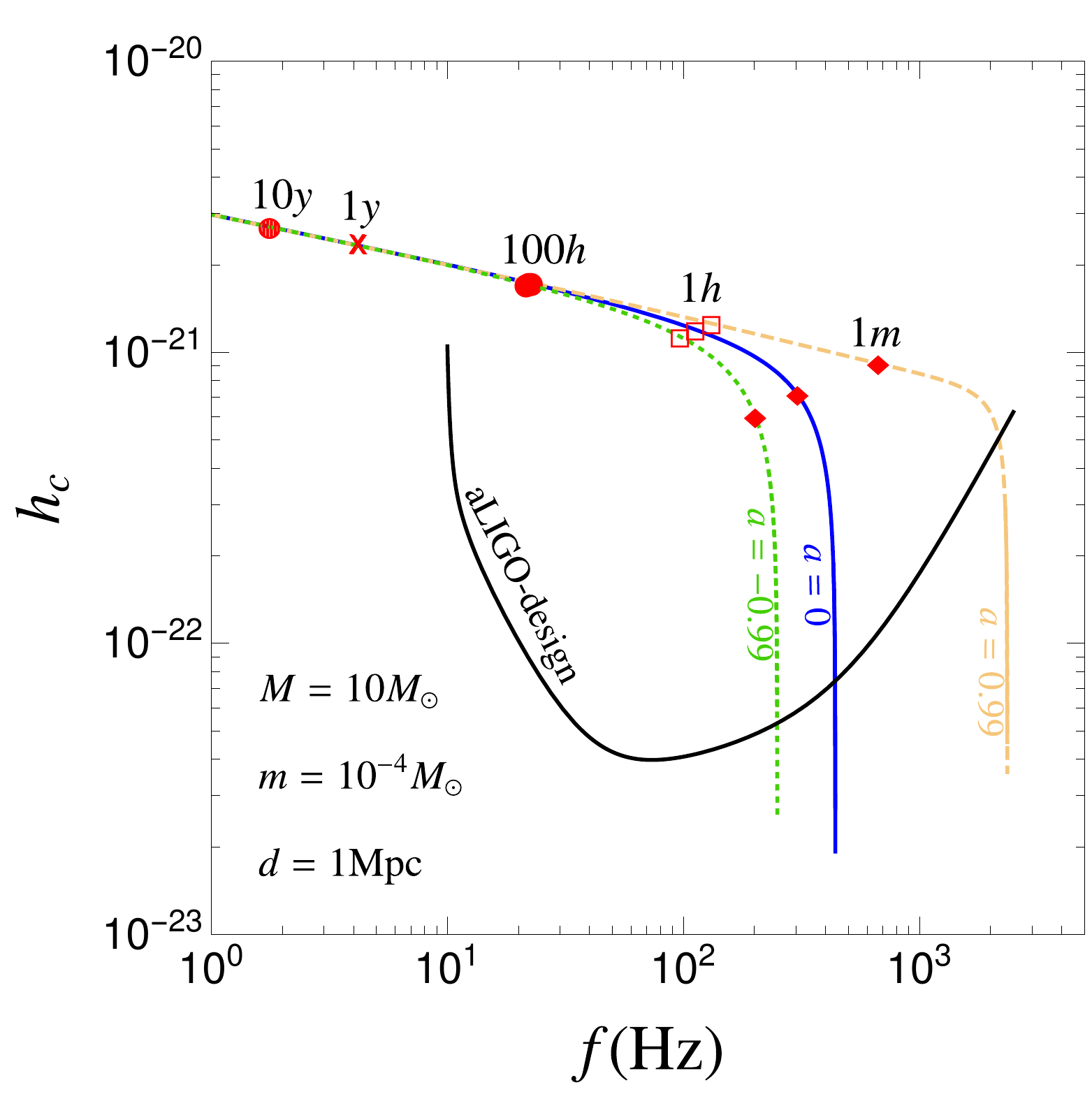}
  \quad
  \caption{
\label{fig:hc}
The dimensionless characteristic strain $h_c$ for a mini-EMRI with mass ratio $10^{-5}$ at a distance of $d$ from the detector, for $a=-0.99,0,0.99$, where
the markers denote remaining times to ISCO: 1 minute, 1 hour, 100 hours, 1 year, and 10 years.
 }
\end{figure}

\noindent{\bfseries Gravitational-Wave Signal Properties.}
The calculation of the \gwh signal from a \miniemri system is similar to that of two approximately equal-mass objects when they are far away from each
other, such that a post-Newtonian treatment is sufficient, which corresponds to the early inspiral stage of the inspiral-merger-ring-down 
waveform~\cite{Ajith:2009bn,Ajith:2007kx}. 
However, as they approach each other, relativistic effects become significant
enough that a fully numerical calculation is necessary, which is a difficult task and still an ongoing effort 
(see \cite{Babak:2017tow,Amaro-Seoane:2007osp,Poisson:2011nh} for reviews). 
While full numerical relativity simulations are advancing toward higher mass ratios (see, e.g., \cite{Lousto:2020tnb,Lousto:2022hoq}), 
the extreme mass ratio of this system makes possible a perturbation theory
of a different kind, which is based on an expansion in the small mass ratio, in the waveform computation (see, e.g., \cite{Poisson:2011nh}).
We use the fully numerical result obtained for quasi-circular evolution of the binary~\cite{Finn:2000sy} based on the Teukolsky formalism~\cite{Teukolsky:1973ha,Sasaki:1981sx}, 
which is consistent~\cite{Caprini:2015zlo} with those obtained based on the numerical or analytical Kludge waveforms~\cite{Babak:2006uv,Chua:2017ujo}.
Fig.~\ref{fig:hc} shows the characteristic strain $h_c(f)$ for the left red vertical line of Fig.~\ref{fig:fISCO} as the blue line, as well as for \miniemri systems with spins of $0.99$ and $-0.99$. We can see that this signal could spend a very long time (relative to detected binary black hole and neutron star mergers) in current and future ground-based 
\gwh detectors' most sensitive frequency bands.


\begin{figure}[t]
\includegraphics[width=0.47\columnwidth]{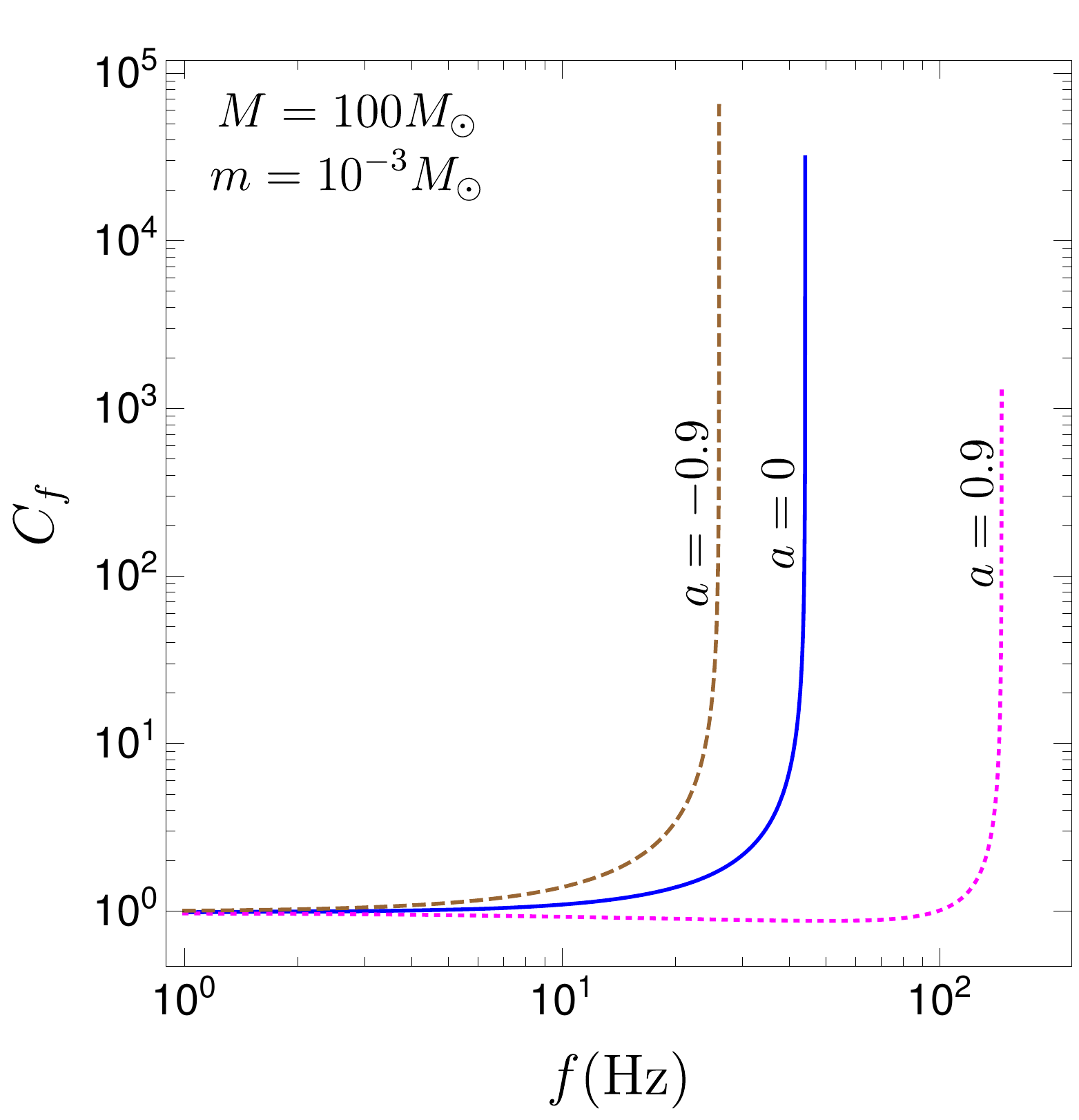}
\quad
\includegraphics[width=0.47\columnwidth]{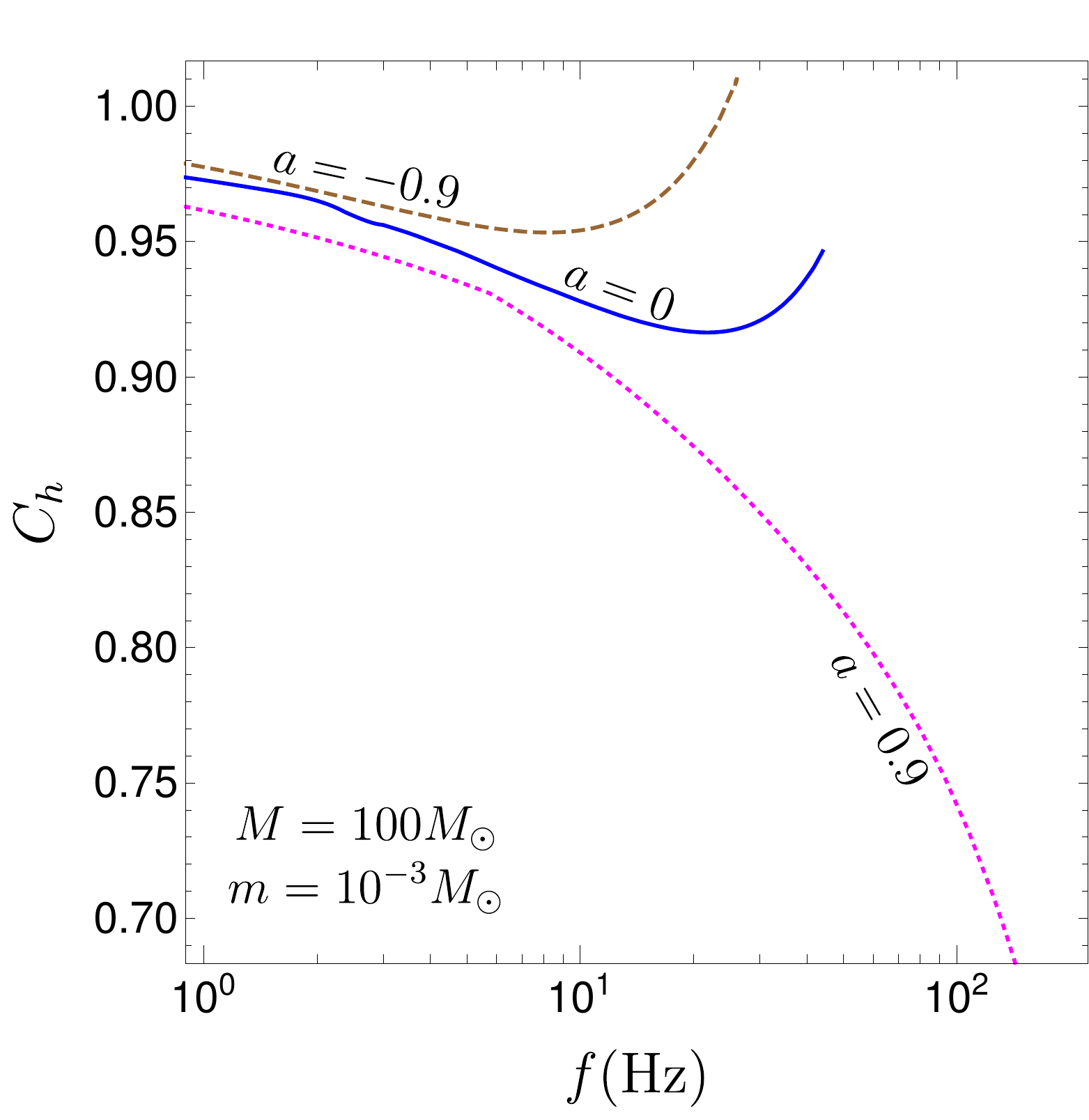}
\caption{\label{fig:dEdf}
The relativistic correction factors for $df/dt$ and $h_0$ as a function of
frequency $f$ for an \miniemri with parameters specified in the figures, for 
three choices of spin $a$.
\label{fig:Cfactors}
}
\end{figure}
The \gwh signal can be divided into two stages with different properties: the early inspiral and the late plunging.
For the inspiral part, the signal evolves with a frequency ``spin-up'' rate $df/dt$:
\begin{eqnarray}
\frac{df}{dt} &=& \frac{96}{5} \pi^{8/3} \left(\frac{G M_c}{c^3}\right)^{5/3} f^{11/3} C_f(a, f) ,
\end{eqnarray}
where $G$ is Newton's gravitational constant and $c$ is the speed of light. 
The factor $C_f$, as shown in the left panel of Fig.~\ref{fig:Cfactors}, captures relativistic effects and asymptotes to $1$ for small $f$, or a large separation. As the ISCO is approached, $C_f$ increases rapidly and diverges corresponding to the second plunging phase.
The \gwh emission, as represented by $h_c$, decreases significantly during this stage, making its detection potentially difficult.

Since $M_c$ is small for subsolar mass searches, the frequency increases slowly during this stage.
At the same time, the \gwh amplitude evolves with a similar relativistic correction factor $C_h$ (as shown in the right panel of Fig.~\ref{fig:Cfactors}):
\begin{eqnarray}
h_0 &=& \frac{4}{d} \left(\frac{G M_c}{c^2}\right)^{5/3} \left(\frac{\pi f}{c}\right)^{2/3} C_h (a, f) .
\end{eqnarray}
Thus, the signal does not behave like a typical
chirp signal that lasts for $\mathcal{O}($seconds), but instead like continuous waves from neutron stars, which LIGO/Virgo are actively searching for using techniques and methodology readily applicable here \cite{Krishnan:2004sv,Astone:2014esa,Miller:2018rbg,Sun:2018hmm}. This means that directed searches~\cite{LIGOScientific:2021quq} for mini-EMRI systems, in which the heavier object could be a known black hole or neutron star,
could be performed, as well as all-sky searches~\cite{LIGOScientific:2022pjk}, in which the heavier mass would be of arbitrary origins. 

Compared with previous searches for compact ECOs, allowing a generic compactness $C$ leads to 
significant changes in the \gw signals, since the ECO might be tidally disrupted before reaching the ISCO, and correspondingly
the \gw
signal will be cut off at this frequency~\cite{Guo:2019sns}. The tidal radius can be estimated by 
equating the gravitational force of the heavier object with the self-gravitating force of the lighter one as $r_{\text{Tidal}}= (m^2 M)^{1/3}/C$, which
then translates into another maximal frequency cut-off $f_{\text{Tidal}}$, implying that the actual 
cut-off frequency will be $min(f_{\text{Tidal}}, f_{\text{ISCO}})$. Additionally, the tidal disruption 
could provide potentially valuable electromagnetic counterpart signals to a \gwh detection of a mini-EMRI, which could reveal nature of
the underlying ECO and help probe new physics.

\begin{figure}[t]
\includegraphics[width=0.8\columnwidth]{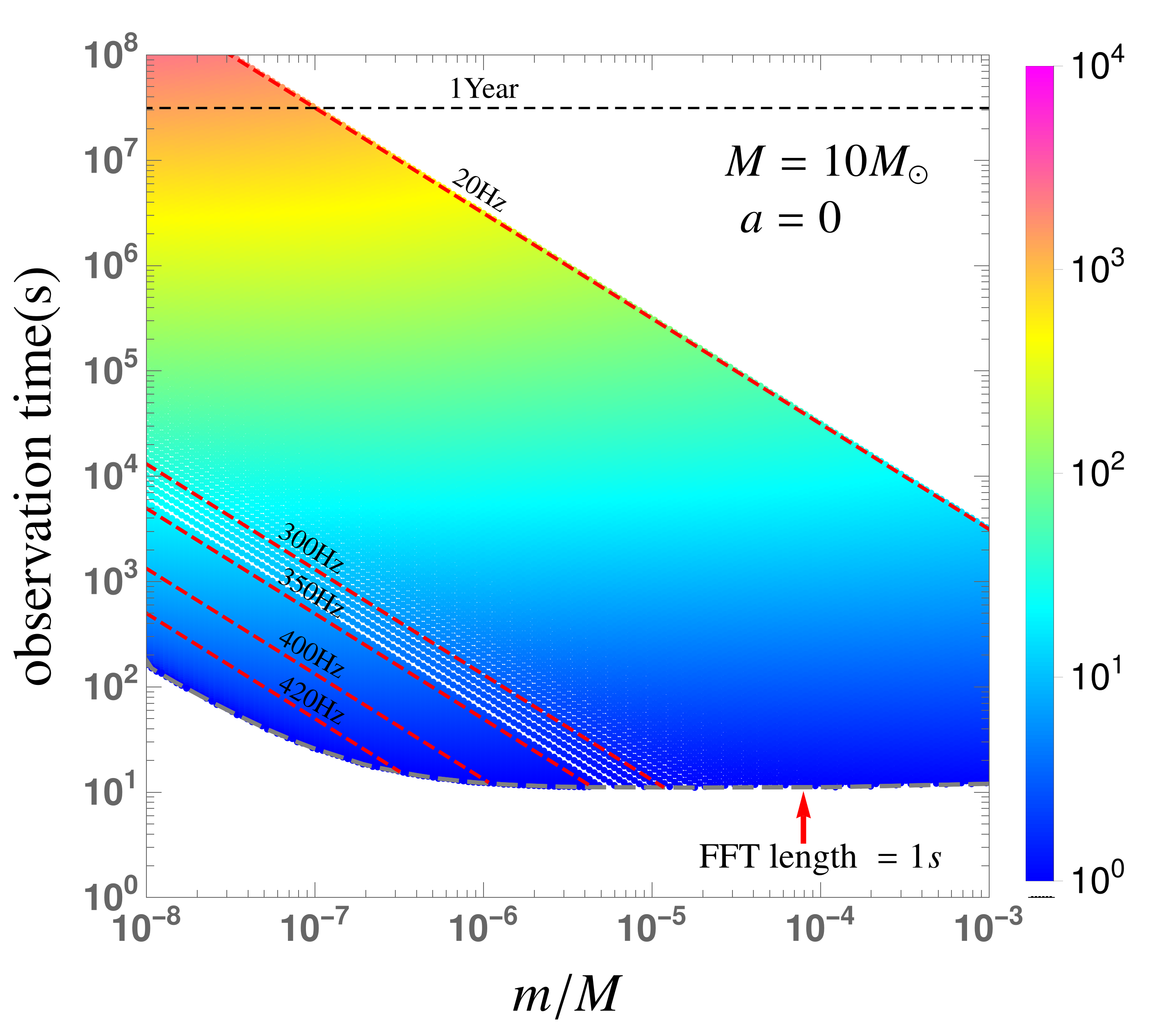}
\caption{The observation time with color coding of FFT length (a minimum of 1 second) as a function of the mass ratio, 
for $M=10M_\odot$, and a 10-Hz band in the analysis, with a starting frequency ranging from 20Hz to $f_{\text{ISCO}}$. 
}
\label{fig:obs_fft_space}
\end{figure}

\noindent{\bfseries Search with the Hough Transform.} We now employ strategies, developed in the context of \cwh searches, that attempt to detect \gws from asymmetrically rotating neutron stars \cite{Riles:2017evm,sieniawska2019continuous}. At their cores, these methods assume that the \gwh frequencies evolve linearly and slowly with time \cite{Astone:2014esa,Krishnan:2004sv,LIGOScientific:2007hnj}. There also exist so-called ``transient'' \cwh methods \cite{Miller:2018rbg,Miller:2019jtp,Sun:2018hmm,Oliver:2018dpt,banagiri2019search,mytidis2015sensitivity}, originally developed to search for remnants of neutron star mergers or supernovae \cite{longpmr}, that track rapid power-law frequency evolutions over time. The inspiral portion of \miniemri systems would follow a power-law frequency evolution over time, but would reduce to a linear one if the mass ratio is small enough \cite{Miller:2020kmv}. In both cases, we can employ the Hough Transform to search for \gwh signals from \miniemris. This method maps points in the time/frequency plane to lines in the frequency/frequency derivative plane of the source \cite{Astone:2014esa,Miller:2018rbg}. While not as sensitive as matched filtering, the Hough Transform has been shown to be robust against noise disturbances, gaps and generally non-stationary noise \cite{Astone:2014esa,Krishnan:2004sv}, three problems that would likely occur over the duration of the \miniemri signal in \gwh data. 

In this method, we break up the data into chunks of durations $\TFFT$ much less than the signal duration $\Tobs$, Fast Fourier transform (FFT) each chunk, and combine the power in each chunk incoherently. The length of each FFT is chosen to confine the signal power in one frequency bin during that FFT, which means that it is primarily a function of the frequency change of the inspiraling system over time \cite{Miller:2020kmv}. The choice of $\TFFT$ significantly affects the sensitivity, or distance reach, of the search. To choose this quantity, for a range of mass ratios, we calculate how quickly the system inspirals, and use that to determine $\TFFT$ and $\Tobs$, within a given 10-Hz band. The choice of a 10-Hz analysis band is arbitrary, but in a real search, an upper-frequency cutoff must be selected, since observing the signal until the ISCO would not result in a good sensitivity, as $\TFFT$ would have to be very small to contain the frequency modulation induced by the large spin-up. Instead, we could optimize the choice of the analysis frequency band, as a function of the mass ratio, as done in the case of equal planetary-mass primordial black hole systems \cite{Miller:2020kmv}, which will be the subject of future work.

Here, we set $\TFFT=1/\sqrt{\dot{f}}$ and evaluate $\dot{f}$ at the highest frequency in the 10-Hz band, which will lead to a conservative estimation of the sensitivity.
We show in Fig.~\ref{fig:obs_fft_space} an example of a \miniemri system with $M=10 M_{\odot}$, $a=0$, in which the y-axis shows the total observation time
of the 10-Hz band and the x-axis shows the mass ratio. For each mass ratio, a set of
observation times is obtained by varying the starting frequency from 20Hz to 
a value near $f_{\text{ISCO}}$, leading to a region on this plot.
The color here denotes the FFT length, whose minimum value is set to be one second, which corresponds to the lower boundary of the colored region. Points with the same starting frequency for different mass ratios fall on a line, as labeled for several in the plot.
Typically, lighter-mass systems at lower frequencies will exhibit more linear and slower frequency evolutions over time than those with higher-mass, higher-frequency counterparts \cite{Miller:2020kmv}. Furthermore, the former systems will inspiral for a lot longer due to having smaller changes in frequency over time than those of the latter systems. 


\begin{figure*}[t]
\includegraphics[width=0.49\textwidth]{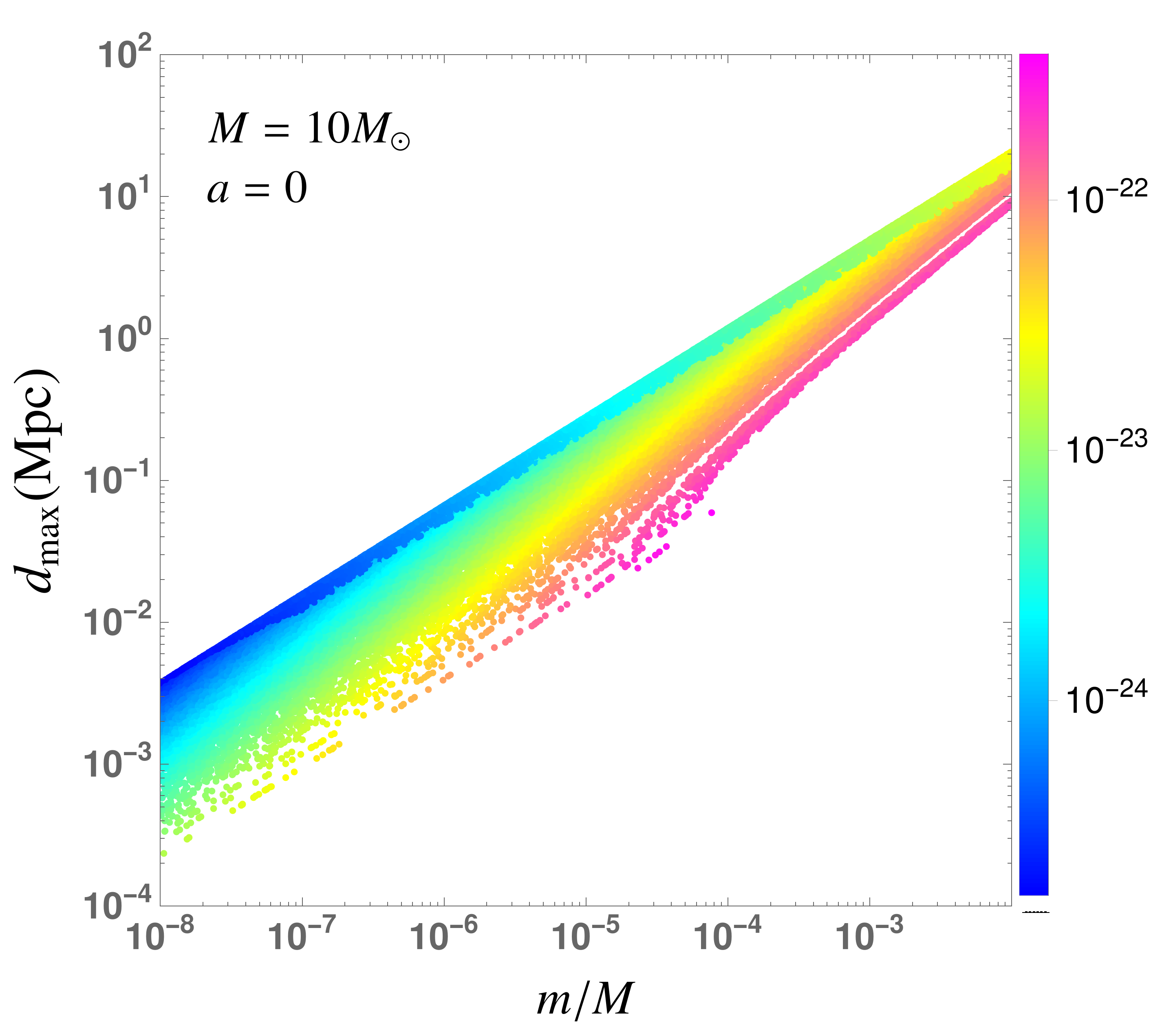}
\quad
  \includegraphics[width=0.435\textwidth]{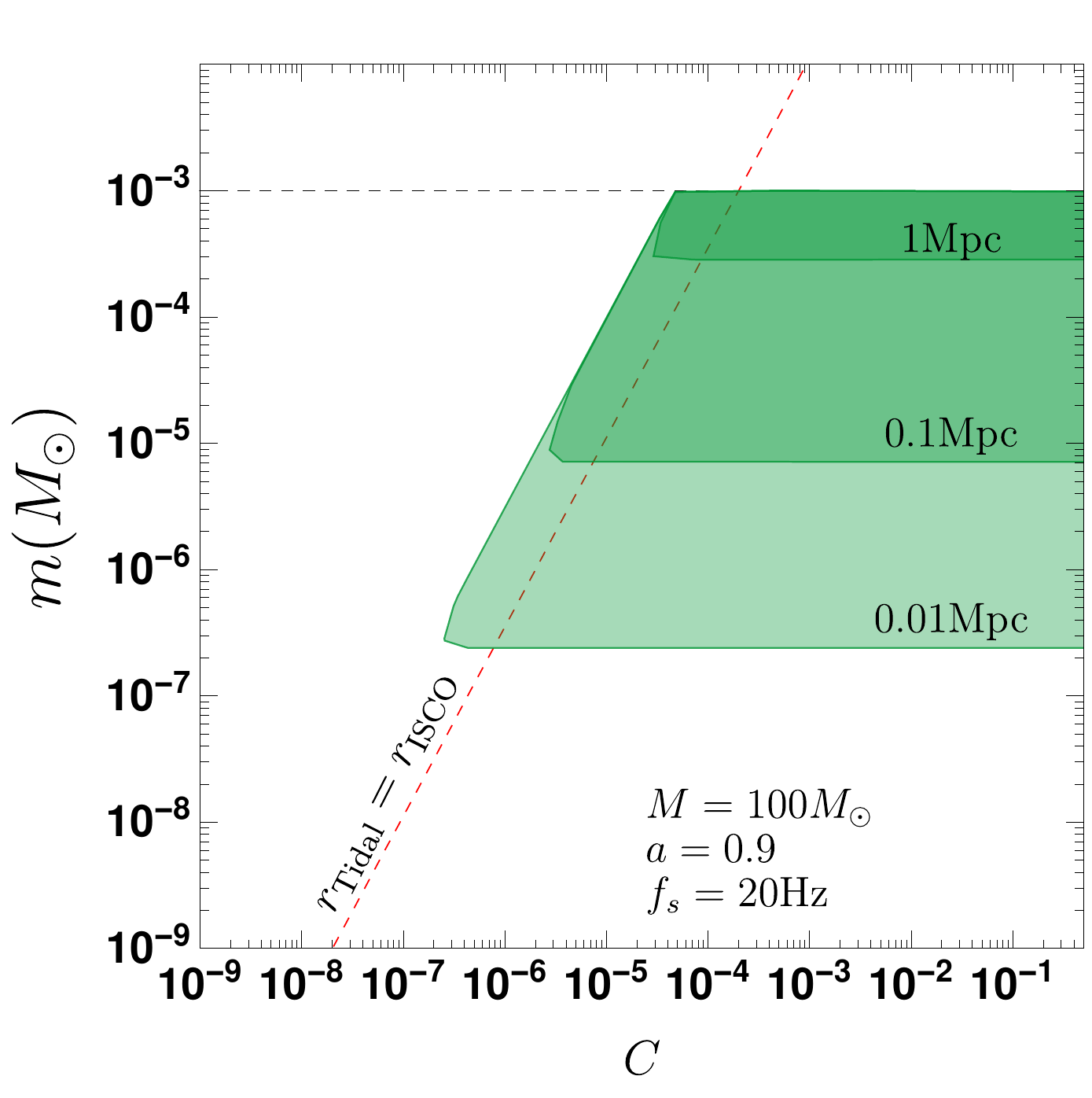}
  \caption{\label{fig:res}
  Left: estimated distance reach at 95\% confidence as a function of mass ratio, with the minimum detectable \gwh amplitude colored. Right: LIGO sensitivity to the \miniemri consisting of a $100 M_{\odot}$ central massive compact object and an ECO specified by its mass (vertical axis) and compactness (horizontal axis), where
  the color-shaded regions denote where the corresponding mini-EMRI system can emit gravitational waves and be detected by LIGO/Virgo, assuming a CR threshold of 5.
  } 
\end{figure*}

To estimate the sensitivity to \miniemri binaries, we employ the method in \cite{Miller:2020kmv} to calculate the minimum detectable \gwh amplitude at a given confidence level, as a function of the analysis coherence time, the signal duration, and particular analysis parameters. This minimum amplitude can be translated into 
a maximum distance reach $d_{\rm max}$
\begin{widetext}
\begin{equation}
d_{\rm max}=0.995\left(\frac{G M_c}{c^2}\right)^{5/3}\left(\frac{\pi}{c}\right)^{2/3} \frac{\TFFT}{\sqrt{\Tobs}}\left(\sum_i \frac{\mathcal{F}^2_i}{S_n(f_i)}\right)^{1/2}\left(\frac{p_0(1-p_0)}{Np^2_1}\right)^{-1/4}\sqrt{\frac{\theta_{\rm thr}}{\left(CR_{\rm thr}-\sqrt{2}\erfc^{-1}(2\Gamma)\right)}},
\label{dmax}
\end{equation}
\end{widetext}
where $\mathcal{F}_i=f_i^{2/3} C_h(a,f_i)$; $i$ is an index that runs over the chunks; $N$ is the number of FFTs in $\Tobs$; $p_0$ is the probability of selecting a noise peak above a threshold; $p_1$ = $e^{-\theta_{\rm thr}}-$  2$e^{-2\theta_{\rm thr}}$ $+e^{-3\theta_{\rm thr}}$, $\theta_{\rm thr}=2.5$ is a threshold on equalized power spectra in the time/frequency map; $\Gamma=0.95$ is the confidence level; $CR_{\text{thr}}=5$ is the 
threshold of the critical ratio in selecting candidates in the frequency-Hough map; and $S_n$ is an estimation of the noise power spectral density of the detector. 



\noindent{\bfseries Results.}
The hypothetical distance reach and minimum detectable amplitude of a search for \miniemri systems are shown in the left panel of
Fig.\ref{fig:res} on the y-axis and in color, respectively, as a function of the mass ratio for a system with $M=10M_\odot$, $a=0$ two compact objects. The starting frequency of the 10-Hz band is varied from 20Hz
to a frequency close to $f_{\text{ISCO}}$. A fixed starting frequency
results in a line in this plot and varying it leads to
the region shown here.
For a fixed mass ratio, increasing the starting frequency from 20Hz leads firstly 
to an increasing distance reach before dropping to lower values, which is different
than in Fig.\ref{fig:obs_fft_space}, and leads to the mixing of the blue and cyan colors. Systems with less extreme mass ratios could be seen farther away, at $\mathcal{O}(10\text{Mpc})$, while those with more extreme mass ratios could only be detected $\mathcal{O}$(kpc-Mpc) away.


The result of allowing the lighter one, with mass $m$, of the \miniemri system to be non-compact, 
with compactness $C$, is shown in the right panel of \ref{fig:res}. Regions on the plane $(m,C)$ that
could be detected are denoted by the green region with the contours for several distances shown. The red dashed line denotes
where the tidal radius coincides with the ISCO radius, while for the region to the left (right) of this line the tidal radius is larger (smaller) than the ISCO radius. As the tidal radius becomes larger and
cuts into the 10Hz band, the \gw signal available to build up the sensitivity is gradually lost, which then leads to the boundary of the
green region.


\noindent{\bfseries Discussion.}
In this \emph{letter}, we propose an innovative idea to search current ground-based \gwh data for \miniemri systems. A detection of such a source would imply a huge paradigm shift in the way that we understand compact objects. The prospects of seeing such a system within our galaxy in the current detector era are promising, based on the sensitivity estimation presented here, and even more possible when Einstein Telescope and Cosmic Explorer come online. We also propose a way to search for such systems with the Hough Transform, that would track the quasi-power-law nature of the inspiraling mini-EMRI system and provide an estimate of the chirp mass.
Our work paves the way for searches for \miniemri systems, and demonstrates the effectiveness of using traditional \cwh methods to probe the existence of \ecos.

\noindent{\bfseries Acknowledgements.} 
This material is based upon work supported by NSF's
LIGO Laboratory which is a major facility fully funded
by the National Science Foundation.
This research has used data obtained from the
Gravitational Wave Open Science Center, a service of
LIGO Laboratory, the LIGO Scientific Collaboration and
the Virgo Collaboration. 
LIGO Laboratory
and Advanced LIGO are funded by the United States
National Science Foundation (NSF) as well as the Science and Technology Facilities Council (STFC) of the
United Kingdom, the Max-Planck-Society (MPS), and
the State of Niedersachsen/Germany for support of the
construction of Advanced LIGO and construction and
operation of the GEO600 detector. Additional support
for Advanced LIGO was provided by the Australian Research Council. Virgo is funded, through the European
Gravitational Observatory (EGO), by the French Centre
National de Recherche Scientifique (CNRS), the Italian
Istituto Nazionale della Fisica Nucleare (INFN) and the
Dutch Nikhef, with contributions by institutions from
Belgium, Germany, Greece, Hungary, Ireland, Japan,
Monaco, Poland, Portugal, and Spain.
We thank Juan Calderon Bustillo, Carlos Lousto, Cristiano Palomba, Kuver Sinha, Yue Zhao, and the LIGO/Virgo/KAGRA continuous-wave group 
for helpful comments and discussions.
HG is supported by the U.S. Department of Energy under Award No. DESC0009959.
ALM is a beneficiary of a FSR Incoming Post-doctoral Fellowship.


\bibliographystyle{utphys}

\providecommand{\href}[2]{#2}\begingroup\raggedright\endgroup
\end{document}